\documentclass[prd,preprintnumbers,showpacs,onecolumn,showkeys,floatfix,nofootinbib,notitlepage]{revtex4-1}
\usepackage{graphicx}
\usepackage{amssymb}
\usepackage{amsmath}
\usepackage{mathtools}
\usepackage{epsfig}
\usepackage{color}
\usepackage{enumerate}
\usepackage{slashed}
\usepackage{hyperref}
\usepackage{epstopdf}

\def\slashchar#1{\setbox0=\hbox{$#1$}
   \dimen0=\wd0 \setbox1=\hbox{/} \dimen1=\wd1
   \ifdim\dimen0\big>\dimen1 \rlap{\hbox to \dimen0{\hfil/\hfil}} #1
   \else  \rlap{\hbox to \dimen1{\hfil$#1$\hfil}} / \fi}

\newcommand{\ud}{\mathrm{d}}
\newcommand{\be}{\begin{equation}}
\newcommand{\ee}{\end{equation}}
\newcommand{\bea}{\begin{eqnarray}}
\newcommand{\eea}{\end{eqnarray}}

\newcommand{\Appendix}[1]%
    {%
     \section{#1}%
      }

\begin{document}

\title{Quantization of Yang-Mills Theory Without the Gribov Ambiguity}

\author{Gao-Liang Zhou}
\email{ zhougl@itp.ac.cn}
\affiliation{College of Science, Xi'an University of Science and Technology, Xi'an 710054, People's Republic of China}
\author{Zheng-Xin Yan}
\affiliation{College of Science, Xi'an University of Science and Technology, Xi'an 710054, People's Republic of China}
\author{Xin Zhang}
\affiliation{College of Science, Xi'an University of Science and Technology, Xi'an 710054, People's Republic of China}




\begin{abstract}
A gauge fixing condition is presented here for non-Abelian gauge theory on the manifold $R\otimes S^{1}\otimes S^{1}\otimes S^{1}$. It is proved that the new gauge fixing condition is continuous and free from the Gribov ambiguity. While perturbative calculations based on the new gauge condition behave like those based on the axial gauge in ultraviolet region,  infrared behaviours of the perturbative series under the new gauge fixing condition are quite nontrivial. The new gauge condition, which reads $n\cdot\partial n\cdot A=0$, may not satisfy the boundary condition $A^{\mu}(\infty)=0$ as required by conventional perturbative calculations for gauge theories on the manifold $S^{4}$. However, such contradiction is not harmful  for the theory considered here.
\end{abstract}


\maketitle

\section{Introduction.}
\label{introduction}

Gauge fixing procedure of non-Abelian gauge theory is a nontrivial issue and hampered by some ambiguities\cite{Gribov-1978,Singer-1978}. The conventional Faddeev-Popov quantization procedure\cite{Faddeev:1967fc} is based on the equation,
\begin{equation}
\int[\mathcal{D}\alpha(x)]\textrm{det} (\frac{\delta(G(A))}{\delta \alpha})\delta(G(A))=1,
\end{equation}
where $\alpha(x)$ represents the parameter of gauge transformation, $G(A)$ represents the gauge fixing function and $G(A)=\partial A$ for the Landau gauge. In \cite{Gribov-1978}, the author shows that the Landau gauge $\partial\cdot A=0$  is not a good gauge fixing condition for non-Abelian gauge theories as it does not intersect with each gauge orbit exactly once. Such ambiguity is termed as Gribov ambiguity in literatures.  In \cite{Singer-1978}, it was proved that there is no continuous gauge fixing condition which is free from the  Gribov ambiguity  for non-Abelian gauge theory  on $3$-sphere($S^{3}$) and $4-$sphere($S^{4}$) once the gauge group is compact.

The Gribov ambiguity is related to the zero eigenvalues(with nontrivial eigenvectors) of the Faddeev-Popov operator\cite{Gribov-1978,vanBaal:1991zw,Sobreiro:2004us,Vandersickel:2012tz}. It seems natural to work in the so-called Gribov region\cite{Gribov-1978,Vandersickel:2012tz},  in which the Faddeev-Popov operator is positive definite. The Gribov region is convex and intersects with each gauge orbit at least once\cite{Zwanziger:1982na,Dell'Antonio:1991xt}. Integral region of the gauge potential is  restricted to the Gribov region through the no pole condition\cite{Gribov-1978,Sobreiro:2004us}, which means that nontrivial poles of propagators of ghosts should vanish in the Gribov region.  Such restriction can also be realized trough  the Gribov-Zwanziger(GZ) action\cite{Zwanziger:1988jt,Zwanziger:1989mf,Zwanziger:1992qr}. Equivalence between these two methods is proved in \cite{Capri:2012wx}. The Gribov region method is extended to general $R_{\xi}$ gauges in \cite{Lavrov:2013boa} through the field dependent BRST transformation\cite{Joglekar:1994tq,Lavrov:2013rla}. The method can also be extended to the maximal Abelian gauge(see, e.g. Refs. \cite{Capri:2006cz,Capri:2010an,Capri:2015pfa}).

Although researches based on the GZ action are interesting and fruitful(see, e.g. Refs.\cite{Dudal:2007cw,Dudal:2008sp,Su:2014rma,Bandyopadhyay:2015wua,Guimaraes:2015vra} ).
There is still Gribov ambiguity even  if one works in the Gribov region. A possible solution to the Gribov problem  is to work in the absolute Landau gauge\cite{Vandersickel:2012tz,vanBaal:1991zw}, which is the set of the absolute minima of the functional
\begin{equation}
\int\ud^{4}x tr[A^{U}_{\mu}(x)A^{U}_{\mu}(x)],
\end{equation}
where $U$ represents an arbitrary gauge transformation. It is, however, difficult to perform analytical calculations in this gauge.
An alternative way is to average over Gribov copies  as in \cite{Serreau:2012cg,Serreau:2013ila}, which avoids
the Neuberger zero problem of the standard Fadeev-Popov quantization procedure.
One may also take an extra constraint introduced in \cite{Pereira:2013aza,Pereira:2014apa} that eliminates infinitesimal Gribov copies  without the geometric approach.

For an algebraic gauge condition like the axial gauge $n\cdot A=0$, the degeneracy is independent of the gauge potential.  It seems that calculations in such gauge are not affected by the Gribov ambiguity. However, such gauge condition  is not continuous for gauge theories  on the manifold $S^{4}$. To see this, we consider the equation,
\begin{equation}
U n\cdot A U^{\dag}+\frac{i}{g}U n\cdot\partial U^{\dag}=0\Rightarrow U^{\dag}(x)\propto P\exp(ig\int_{-\infty}^{0}n\cdot A(x+sn^{\mu})).
\end{equation}
It is impossible to choose the proportional function in above equation so that $U(x)$ takes unique value at infinity.  This is in contradiction with the continuity of gauge transformations as the infinity is an ordinary point on $S^{4}$. There is another famous algebraic gauge termed as the space-like planar gauge\cite{Kummer:1976hv,Dokshitzer:1978hw,Bassetto:1983rq,Leibbrandt:1987qv}, in which the gauge fixing term reads,
\begin{equation}
\mathcal{L}_{\textrm{fix}}\equiv -\frac{1}{n^{2}} tr[n\cdot A\partial^{2}n\cdot A],
\end{equation}
where $n^{\mu}$ is a space like vector. The space-like planar gauge is free from the Gribov ambiguity\cite{Bassetto:1983rq} and not continuous for gauge theories  on $S^{4}$.

In this paper, we consider non-Abelian gauge theory on the $3+1$ dimensional manifold $R\otimes S^{1}\otimes S^{1}\otimes S^{1}$. Topological properties of the manifold are interesting and may be related to confinement of quarks as displayed in \cite{tHooft:1981sps}.   Gauge potentials on the manifold $R\otimes S^{1}\otimes S^{1}\otimes S^{1}$ satisfy the  periodic boundary conditions,
\begin{eqnarray}
\label{boundary condition}
A^{\mu}(t,x_{1}+L_{1},x_{2},x_{3})&=&A^{\mu}(t,x_{1},x_{2},x_{3})
\nonumber\\
A^{\mu}(t,x_{1},x_{2}+L_{2},x_{3})&=&A^{\mu}(t,x_{1},x_{2},x_{3})
\nonumber\\
A^{\mu}(t,x_{1},x_{2},x_{3}+L_{3})&=&A^{\mu}(t,x_{1},x_{2},x_{3}),
\end{eqnarray}
where $L_{i}$($i=1,2,3$) are large constants. It is hard to maintain Lorentz invariance in theories on the manifold. We do not consider such defect here. We will show that the gauge condition,
\begin{equation}
\label{gaugex}
n\cdot\partial n\cdot A=0
\end{equation}
is continuous and free from the Gribov ambiguity for gauge theories on the manifold $R\otimes S^{1}\otimes S^{1}\otimes S^{1}$, where $n^{\mu}$ represents directional vectors along $x_{i}$-axis($i=1,2,3$).
We can rewrite the gauge condition in momentum space, which reads,
\begin{equation}
\label{gaugek}
n\cdot A(k)=0(\text{for $n\cdot k\ne 0$}).
\end{equation}
 We see that the gauge fixing condition is equivalent to the axial gauge for $n\cdot k\ne 0$.

The paper is organized as follows. In Sec.\ref{R3torus}, we describe gauge theory on $R\otimes S^{1}\otimes S^{1}\otimes S^{1}$ briefly. In Sec.\ref{quantization}, we consider non-Abelian gauge theory on $R\otimes S^{1}\otimes S^{1}\otimes S^{1}$ and present the proof that the gauge condition (\ref{gaugex}) is continuous and free from the Gribov ambiguity. In Sec.\ref{perturbationserie}, we discuss propagators of gluons under the new gauge fixing condition. Our conclusions and some discussions are presented in Sec.\ref{conc}.

\section{Gauge Theories on $R\otimes S^{1}\otimes S^{1}\otimes S^{1}$}
\label{R3torus}

In this section, we describe gauge theories on the manifold $R\otimes S^{1}\otimes S^{1}\otimes S^{1}$. The manifold $R\otimes S^{1}\otimes S^{1}\otimes S^{1}$ can be obtained from the Minkowski space through the identification
\begin{equation}
(t,x_{1},x_{2},x_{3})\sim (t,x_{1}+L_{1},x_{2},x_{3})\sim (t,x_{1},x_{2}+L_{2},x_{3})\sim (t,x_{1},x_{2},x_{3}+L_{3}),
\end{equation}
where $L_{i}$($i=1,2,3$) are  large constants.
 We  take the following periodic boundary conditions,
\begin{eqnarray}
\label{periocond}
A^{\mu}(t,x_{1}+L_{1},x_{2},x_{3})&=&A^{\mu}(t,x_{1},x_{2},x_{3})
\nonumber\\\
A^{\mu}(t,x_{1},x_{2}+L_{2},x_{3})&=&A^{\mu}(t,x_{1},x_{2},x_{3})
\nonumber\\
A^{\mu}(t,x_{1},x_{2},x_{3}+L_{3})&=&A^{\mu}(t,x_{1},x_{2},x_{3})
\end{eqnarray}
in this paper. Effects of the center vortexes like those shown in\cite{'tHooft:1977hy} are not considered here. We require that
\begin{eqnarray}
\label{perioga}
U(t,x_{1}+L_{1},x_{2},x_{3})&=&U(t,x_{1},x_{2},x_{3})
\nonumber\\\
U(t,x_{1},x_{2}+L_{2},x_{3})&=&U(t,x_{1},x_{2},x_{3})
\nonumber\\
U(t,x_{1},x_{2},x_{3}+L_{3})&=&U(t,x_{1},x_{2},x_{3}),
\end{eqnarray}
for continuous gauge transformation on the manifold $R\otimes S^{1}\otimes S^{1}\otimes S^{1}$.

Quantum field theories on $R\otimes S^{1}\otimes S^{1}\otimes S^{1}$ are quite similar to quantum mechanics in the box normalization scheme. In such scheme the momentum operator $-i\vec{\nabla}$ is a Hermitian operator as surface terms vanish according to periodic boundary conditions. For quantum field theories on $R\otimes S^{1}\otimes S^{1}\otimes S^{1}$, the surface terms also vanish according to periodic conditions (\ref{periocond}). Thus the operator $-i\vec{\nabla} A^{\mu}$ is Hermitian. We can get perturbative series similar to those in quantum field theory on $S^{4}$.

We should emphasize here that the manifold $R\otimes S^{1}\otimes S^{1}\otimes S^{1}$ is not Lorentz invariant, which seems troublesome.  The manifold $R\otimes S^{1}\otimes S^{1}\otimes S^{1}$ looks like the Minkowski space locally. It seems to us that the Lorentz invariance can be restored for local quantities in the limit $L_{i}\to\infty$($i=1,2,3$). In fact, Feynman rules of quantum theories on the manifold $R\otimes S^{1}\otimes S^{1}\otimes S^{1}$ are similar to those on the manifold $R^{4}$ except for that momenta of particles take discrete values for the theory considered here. For the case that $L_{i}\to\infty$($i=1,2,3$), summations over discrete momenta values tend to integrals over the momenta space once such integrals are not affected by ultraviolet divergences or mass singularities. In perturbative calculations, ultraviolet divergences are absorbed into physical constants through renormalization  procedures. Mass singularities are harmless for local quantities once the summation over all possible initial and final states have been performed according to the famous  Kinoshita-Lee-Nauenberg(KLN) theorem.\cite{KLN:1962,KLN:1964}. As a result, we simply assume that the Lorentz invariance can be restored in the limit $L_{i}\to\infty$($i=1,2,3$) for local quantities which are multiplicative renormalized and infrared safe. Renormalization properties and KLN cancellations of theories on the manifold $R\otimes S^{1}\otimes S^{1}\otimes S^{1}$ are not considered here.

To explain what happens on the manifold $R\otimes S^{1}\otimes S^{1}\otimes S^{1}$, we consider a gauge theory of which the gauge group is $U(1)$. Although such gauge theory is free from the Gribov ambiguity in Landau gauge, it is convenient to take this theory as an example to show that the gauge condition
\begin{equation}
n\cdot \partial n\cdot A=0
\end{equation}
is a continuous gauge on $R\otimes S^{1}\otimes S^{1}\otimes S^{1}$ and free from the Gribov ambiguity, where $n^{\mu}$ is the directional vector along $x_{i}$-axis($i=1,2,3$). According  to the  boundary conditions (\ref{boundary condition}), we can write $n\cdot A(x)$ as:
\begin{equation}
n\cdot A(x)=\sum_{m}e^{\textrm{i}2\pi m \frac{n\cdot x}{n\cdot L}}f_{m}(x_{T}),
\end{equation}
where $L^{\mu}=(0,L_{1},L_{2},L_{3})$ and $x_{T}$ is defined as
\begin{equation}
x_{T}^{\mu}\equiv x^{\mu}-\frac{n\cdot x}{n^{2}}n^{\mu}.
\end{equation}
A continuous gauge transformation $U(x)=\exp (\textrm{i}\phi(x))$  should also satisfy the boundary conditions (\ref{boundary condition}). We thus have:
\begin{equation}
\phi(x)=2\pi N \frac{n\cdot x}{n\cdot L}+\sum_{m\ne 0 }e^{\textrm{i}2\pi m \frac{n\cdot x}{n\cdot L}}g_{m}(x_{T}),
\end{equation}
where $N$ is an arbitrary integer. Under the gauge transformation $U(x)$, we have
\begin{equation}
n\cdot A^{U}(x)=f_{0}(x_{T})+\frac{2\pi N}{n\cdot L}
+\sum_{m\ne 0}e^{\mathrm{i}2\pi m \frac{n\cdot x}{n\cdot L}}(f_{m}+\frac{2\pi m\textrm{i}}{n\cdot L}g_{m})(x_{T}).
\end{equation}
We can choose suitable $\phi(x)$ so that
\begin{equation}
n\cdot \partial n\cdot A^{U}(x)=0.
\end{equation}
Degeneracy of the gauge condition originates from the arbitrary integer $N$, which is independent of $A^{\mu}(x)$. We see that the gauge condition do eliminate the Gribov ambiguity in this case. For non-Abelian gauge theory, however, the situation is more complicated. We will show that the gauge condition  $n\cdot \partial n\cdot A(x)=0$ is continuous and free from the Gribov ambiguity except for configurations of which the integral measure is $0$.

\section{Quantization of non-Abelian Gauge Theory on $R\otimes S^{1}\otimes S^{1}\otimes S^{1}$ }
\label{quantization}

In this section we consider the quantization of non-Abelian gauge field theory on  $R\otimes S^{1}\otimes S^{1}\otimes S^{1}$.
We consider quantum theory of gauge fields without scalar particles, fermions or massive vector particles here. Lagrangian density of gauge fields can be written as,
\begin{equation}
\mathcal{L}(x)=-\frac{1}{2} tr [G^{\mu\nu }G_{\mu\nu}](x)
\end{equation}
, where $G^{\mu\nu}$ is the gauge field strength tensor,
\begin{equation}
G^{\mu\nu}=\frac{i}{g}[\partial^{\mu}-ig A^{\mu},\partial^{\nu}-igA^{\nu}]
\end{equation}
with $g$ the coupling constant. For a gauge invariant operator $\mathcal{O}(A)$, which is the functional of the gauge field $A^{\mu}(x)$, we have,
\begin{equation}
\big<T\{\mathcal{O}\}\big>=\frac{\int[\mathcal{D}A^{\mu}(x)]\mathcal{O}e^{\mathrm{i}\int\ud^{4}x\mathcal{L}(x)}}
{\int[\mathcal{D}A^{\mu}(x)]e^{\mathrm{i}\int\ud^{4}x\mathcal{L}(x)}},
\end{equation}
where $T$ is the time ordering operator.

Without loss of generality, we choose a special vector $n^{\mu}$ in following texts, where $n^{\mu}$ is defined as
\begin{equation}
n^{\mu}=(0,0,0,1).
\end{equation}

\subsection{The continuous gauge $n\cdot\partial n\cdot A(x)=0$ on $R\otimes S^{1}\otimes S^{1}\otimes S^{1}$}

In this subsection, we prove that the gauge condition $n\cdot\partial n\cdot A(x)=0$ is a  continuous gauge on $R\otimes S^{1}\otimes S^{1}\otimes S^{1}$.
We can write $n\cdot A$ as
\begin{equation}
n\cdot A(x)=\frac{i}{g}Vn\cdot\partial V^{\dag},\quad V=\mathcal{P}\exp (\mathrm{i}g\int_{0}^{x_{3}}\ud z  n\cdot A(x_{0},x_{1},x_{2},z))
\end{equation}
, where $\mathcal{P}$ is the path ordering operator. $V(x)$ may not fulfill the periodic boundary conditions (\ref{perioga}). Thus $V(x)$ is not necessarily to be a continuous gauge transformation on $R\otimes S^{1}\otimes S^{1}\otimes S^{1}$.  We have,
\begin{eqnarray}
\label{bdnong}
V(x_{0},x_{1},x_{2},x_{3}+L_{3})&=&\mathcal{P}\exp (\mathrm{i}g\int_{0}^{x_{3}+L_{3}}\ud z  n\cdot A(x_{0},x_{1},x_{2},z))
\nonumber\\
&=&\mathcal{P}\exp (\mathrm{i}g\int_{L_{3}}^{x_{3}+L_{3}}\ud z  n\cdot A(x_{0},x_{1},x_{2},z))
\mathcal{P}\exp (\mathrm{i}g\int_{0}^{L_{3}}\ud z  n\cdot A(x_{0},x_{1},x_{2},z))
\nonumber\\
&=&V(x)\mathcal{P}\exp (\mathrm{i}g\int_{0}^{L_{3}}\ud z  n\cdot A(x_{0},x_{1},x_{2},z)).
\end{eqnarray}

We then consider the matrix $\mathcal{P}\exp (\mathrm{i}g\int_{0}^{x_{3}}\ud z  n\cdot A(x_{0},x_{1},x_{2},z))$, which can be diagonalized through a unitary transformation.   We have,
\begin{equation}
\mathcal{P}\exp (\mathrm{i}g\int_{0}^{x_{3}}\ud z  n\cdot A(x_{0},x_{1},x_{2},z))=U_{A}e^{\mathrm{i}\phi_{\lambda}}U_{A}^{\dag}(x)
=e^{\mathrm{i}U_{A}\phi_{\lambda}U_{A}^{\dag}}(x),
\end{equation}
where $\phi_{\lambda}$ is a diagonal Hermiatian  matrix  and $U_{A}$ is a unitary matrix. We notice that
\begin{eqnarray}
n\cdot A(x_{0},x_{1}+L_{1},x_{2},z)&=&n\cdot A(x_{0},x_{1},x_{2},z)
\nonumber\\
n\cdot A(x_{0},x_{1},x_{2}+L_{2},z)&=&n\cdot A(x_{0},x_{1},x_{2},z)
\end{eqnarray}
and conclude that the same periodic boundary conditions are satisfied by eigenvalues of the matrix $\mathcal{P}\exp (\mathrm{i}g\int_{0}^{L_{3}}\ud z  n\cdot A(x_{0},x_{1},x_{2},z))$. We have,
\begin{eqnarray}
\label{pebdtr}
e^{\mathrm{i}\phi_{\lambda}}(x_{0},x_{1}+L_{1},x_{2},x_{3})&=&e^{\mathrm{i}\phi_{\lambda}}(x_{0},x_{1},x_{2},x_{3})
\nonumber\\
e^{\mathrm{i}\phi_{\lambda}}(x_{0},x_{1},x_{2}+L_{2},x_{3})&=&e^{\mathrm{i}\phi_{\lambda}}(x_{0},x_{1},x_{2},x_{3}).
\end{eqnarray}
Eigenvalues of the matrixes
\begin{eqnarray}
\phi_{1}(x_{0},x_{1},x_{2},x_{3})&\equiv&\phi_{\lambda}(x_{0},x_{1}+L_{1},x_{2},x_{3})-\phi_{\lambda}(x_{0},x_{1},x_{2},x_{3})
\nonumber\\
\phi_{2}(x_{0},x_{1},x_{2},x_{3})&\equiv&\phi_{\lambda}(x_{0},x_{1},x_{2}+L_{2},x_{3})-\phi_{\lambda}(x_{0},x_{1},x_{2},x_{3})
\end{eqnarray}
equal to $2\pi n$ according to (\ref{pebdtr}), where $n$ represents  arbitrary integers. These integers are independent of coordinates $x_{i}$($i=0,1,2,3$) according to continuity of gauge potentials. We notice that
\begin{equation}
\phi_{\lambda}(x_{0},x_{1},x_{2},0)=0,
\end{equation}
and have
\begin{eqnarray}
\phi_{\lambda}(x_{0},x_{1}+L_{1},x_{2},x_{3})&=&\phi_{\lambda}(x_{0},x_{1},x_{2},x_{3})
\nonumber\\
\phi_{\lambda}(x_{0},x_{1},x_{2}+L_{2},x_{3}&=&\phi_{\lambda}(x_{0},x_{1},x_{2},x_{3}).
\end{eqnarray}
In addition, the periodic boundary conditions
\begin{eqnarray}
U_{A}e^{\mathrm{i}\phi_{\lambda}}U_{A}^{\dag}(x_{0},x_{1}+L_{1},x_{2},x_{3})
&=&U_{A}e^{\mathrm{i}\phi_{\lambda}}U_{A}^{\dag}(x_{0},x_{1},x_{2},x_{3})
\nonumber\\
U_{A}e^{\mathrm{i}\phi_{\lambda}}U_{A}^{\dag}(x_{0},x_{1},x_{2}+L_{2},x_{3})
&=&U_{A}e^{\mathrm{i}\phi_{\lambda}}U_{A}^{\dag}(x_{0},x_{1},x_{2},x_{3}).
\end{eqnarray}
require that
\begin{eqnarray}
[U_{A}^{\dag}(x_{0},x_{1},x_{2},x_{3})U_{A}(x_{0},x_{1}+L_{1},x_{2},x_{3}),e^{\mathrm{i}\phi_{\lambda}}(x_{0},x_{1},x_{2},x_{3})]&=&0,
\nonumber
\end{eqnarray}
\begin{eqnarray}
[U_{A}^{\dag}(x_{0},x_{1},x_{2},x_{3})U_{A}(x_{0},x_{1},x_{2}+L_{2},x_{3}),e^{\mathrm{i}\phi_{\lambda}}(x_{0},x_{1},x_{2},x_{3})]&=&0.
\end{eqnarray}
For the case that eigenvalues of the matrix $e^{\mathrm{i}\phi_{\lambda}}(x_{0},x_{1},x_{2})$ are not degenerate at some points, the matrixes $U_{A}^{\dag}(x_{0},x_{1},x_{2})U_{A}(x_{0},x_{1}+L_{1},x_{2},x_{3})$ and $U_{A}^{\dag}(x_{0},x_{1},x_{2})U_{A}(x_{0},x_{1},x_{2}+L_{2},x_{3})$ are diagonal at these points. We have
\begin{eqnarray}
U_{A}\phi_{\lambda}U_{A}^{\dag}(x_{0},x_{1}+L_{1},x_{2},x_{3})&=&U_{A}\phi_{\lambda}U_{A}^{\dag}(x_{0},x_{1},x_{2},x_{3})
\nonumber\\
U_{A}\phi_{\lambda}U_{A}^{\dag}(x_{0},x_{1},x_{2}+L_{2},x_{3})&=&U_{A}\phi_{\lambda}U_{A}^{\dag}(x_{0},x_{1},x_{2},x_{3})
\end{eqnarray}
at these points. According to the periodic boundary conditions of the matrix
$e^{\mathrm{i}U_{A}\phi_{\lambda}U_{A}^{\dag}}(x_{0},x_{1},x_{2},x_{3})$ and continuity of gauge potentials, we have
\begin{eqnarray}
U_{A}\phi_{\lambda}U_{A}^{\dag}(x_{0},x_{1}+L_{1},x_{2},x_{3})&=&U_{A}\phi_{\lambda}U_{A}^{\dag}(x_{0},x_{1},x_{2},x_{3})
\nonumber\\
U_{A}\phi_{\lambda}U_{A}^{\dag}(x_{0},x_{1},x_{2}+L_{2},x_{3})&=&U_{A}\phi_{\lambda}U_{A}^{\dag}(x_{0},x_{1},x_{2},x_{3})
\end{eqnarray}
at all points once eigenvalues of the matrix $e^{\mathrm{i}\phi_{\lambda}}(x)$ are not degenerate at some points.

We then consider the case that  eigenvalues of the matrix $e^{\mathrm{i}\phi_{\lambda}}(x)$ are degenerate at all points. In this case the matrix $\phi_{\lambda}(x)$ can be decomposed into the form
\begin{equation}
\phi_{\lambda}(x)= \phi_{\lambda;1}(x)+\phi_{\lambda;2}(x),
\end{equation}
where $\phi_{\lambda;1}(x)$ is a Hermitian matrix with degenerate eigenvalues and $\phi_{\lambda;2}(x)$ is a Hermitian matrix which satisfy the  equation
\begin{equation}
\exp(\phi_{\lambda;2}(x))=1.
\end{equation}
We require that eigenvectors of $\phi_{\lambda;1}(x)$ are the same as those of $e^{\mathrm{i}\phi_{\lambda}}(x)$. To explain our decomposition clearly, we give an example here. We consider a continues diagonal unitary matrixes of which eigenvalues are degenerate at all points,
\begin{equation}
e^{\mathrm{i}\phi_{\lambda}}(x)=\left(
\begin{array}{cccccc}
e^{i\lambda} &0       & \ldots        & \ldots           &\ldots            & 0
\\
\vdots       & \vdots & \ldots        & \ldots           &\ldots            & \vdots
\\
0            &\ldots  & e^{i\lambda}  &\ldots            &\ldots            & 0
\\
0            &\ldots  & \ldots        & e^{i\lambda_{1}} &\ldots            & 0
\\
\vdots       & \vdots & \ldots        &\ldots            &\ldots            & \vdots
\\
0            &\ldots  & \ldots        & \ldots           &\ldots            & e^{i\lambda_{n}}
\end{array}
\right)
(x).
\end{equation}
$\phi_{\lambda;1}$ and $\phi_{\lambda;2}$ can be defined as
\begin{equation}
\phi_{\lambda;1}(x)=\left(
\begin{array}{cccccc}
\lambda      &0       & \ldots        & \ldots          &\ldots            & 0
\\
\vdots       & \vdots & \ldots        & \ldots          &\ldots            & \vdots
\\
0            &\ldots  & \lambda       &\ldots           &\ldots            & 0
\\
0            &\ldots  & \ldots        & \lambda_{1}     &\ldots            & 0
\\
\vdots       & \vdots & \ldots        &\ldots           &\ldots            & \vdots
\\
0            &\ldots  & \ldots        & \ldots          &\ldots            & \lambda_{n}
\end{array}
\right),\quad
\phi_{\lambda;2}(x)=\left(
\begin{array}{cccccc}
2\pi N_{1}   &0       & \ldots        & \ldots          &\ldots            & 0
\\
\vdots       & \vdots & \ldots        & \ldots          &\ldots            & \vdots
\\
0            &\ldots  & 2\pi N_{m}    &\ldots           &\ldots            & 0
\\
0            &\ldots  & \ldots        & 0               &\ldots            & 0
\\
\vdots       & \vdots & \ldots        &\ldots           &\ldots            & \vdots
\\
0            &\ldots  & \ldots        & \ldots          &\ldots            & 0
\end{array}
\right),
\end{equation}
where $N_{i}$($i=1,\ldots, m$) are integers. $N_{i}$($i=1,\ldots, m$) are independent of coordinates $x_{j}$($j=0,1,2,3$) according to  continuity of gauge potentials.  We notice that eigenvectors of $\phi_{\lambda;1}(x)$ are the same as those of $e^{\mathrm{i}\phi_{\lambda}}(x)$. Thus matrixes that commute with $e^{\mathrm{i}\phi_{\lambda}}(x)$ should also commute with $\phi_{\lambda;1}(x)$. We have
\begin{eqnarray}
U_{A}\phi_{\lambda;1}U_{A}^{\dag}(x_{0},x_{1}+L_{1},x_{2},x_{3})&=&U_{A}\phi_{\lambda;1}U_{A}^{\dag}(x_{0},x_{1},x_{2},x_{3})
\nonumber\\
U_{A}\phi_{\lambda;1}U_{A}^{\dag}(x_{0},x_{1},x_{2}+L_{2},x_{3})&=&U_{A}\phi_{\lambda;1}U_{A}^{\dag}(x_{0},x_{1},x_{2},x_{3}).
\end{eqnarray}
We consider the traceless Hermitian matrix $\phi(x)$
\begin{equation}
\phi(x)\equiv U_{A}\phi_{\lambda;1}U_{A}^{\dag}(x)+ U_{A} \left(
\begin{array}{cccccc}
2\pi \bar{N}   &0       & \ldots        & \ldots          &\ldots            & 0
\\
\vdots       & \vdots & \ldots        & \ldots          &\ldots            & \vdots
\\
0            &\ldots  & 2\pi\bar{N}    &\ldots           &\ldots            & 0
\\
0            &\ldots  & \ldots        & -2\pi m\bar{N}-tr[\phi_{\lambda;1}]               &\ldots            & 0
\\
\vdots       & \vdots & \ldots        &\ldots           &\ldots            & \vdots
\\
0            &\ldots  & \ldots        & \ldots          &\ldots            & 0
\end{array}
\right)U_{A}^{\dag}(x)
\end{equation}
and have
\begin{eqnarray}
\phi(x_{0},x_{1}+L_{1},x_{2},x_{3})&=&\phi(x_{0},x_{1},x_{2},x_{3})
\nonumber\\
\phi(x_{0},x_{1},x_{2}+L_{2},x_{3})&=&\phi(x_{0},x_{1},x_{2},x_{3}),
\end{eqnarray}
where $\bar{N}$ is an arbitrary integer and the trace of $\phi_{\lambda;1}(x)$ is calculated in color space.
In addition, we notice that
\begin{equation}
e^{\mathrm{i}U_{A}\phi_{\lambda}U_{A}^{\dag}}(x)=e^{\mathrm{i}U_{A}\phi_{\lambda;1}U_{A}^{\dag}}(x)
=e^{\mathrm{i}\phi}(x).
\end{equation}
and conclude that one can always choose a continues traceless Hermitian matrix $\phi(x)$ so that
\begin{eqnarray}
\phi(x_{0},x_{1}+L_{1},x_{2},x_{3})&=&\phi(x_{0},x_{1},x_{2},x_{3})
\nonumber\\
\phi(x_{0},x_{1},x_{2}+L_{2},x_{3})&=&\phi(x_{0},x_{1},x_{2},x_{3}),
\end{eqnarray}
\begin{equation}
e^{\mathrm{i}\phi(x)}=\mathcal{P}\exp (\mathrm{i}g\int_{0}^{x_{3}}\ud z  n\cdot A(x_{0},x_{1},x_{2},z)),
\end{equation}
for the case that  eigenvalues of the matrix $e^{\mathrm{i}\phi_{\lambda}}(x)$ are degenerate at all points.

According to above analyses, we see that
one can always choose a continuous traceless Hermitian matrix $\phi(x)$ so that
\begin{eqnarray}
\phi(x_{0},x_{1}+L_{1},x_{2},x_{3})&=&\phi(x_{0},x_{1},x_{2},x_{3})
\nonumber\\
\phi(x_{0},x_{1},x_{2}+L_{2},x_{3})&=&\phi(x_{0},x_{1},x_{2},x_{3}),
\end{eqnarray}
\begin{equation}
e^{\mathrm{i}\phi(x)}=\mathcal{P}\exp (\mathrm{i}g\int_{0}^{x_{3}}\ud z  n\cdot A(x_{0},x_{1},x_{2},z)).
\end{equation}
It is convenient to bring in the unitary operator $\widetilde{V}(x)$:
\begin{equation}
\widetilde{V}(x)\equiv  e^{\mathrm{i}\frac{x_{3}}{L_{3}}\phi(x_{0},x_{1},x_{2},L_{3})} V^{\dag}(x) .
\end{equation}
According to (\ref{periocond}) and (\ref{bdnong}), we have:
\begin{eqnarray}
\widetilde{V}(t,x_{1}+L_{1},x_{2},x_{3})&=&\widetilde{V}(t,x_{1},x_{2},x_{3})
\nonumber\\\
\widetilde{V}(t,x_{1},x_{2}+L_{2},x_{3})&=&\widetilde{V}(t,x_{1},x_{2},x_{3})
\nonumber\\
\widetilde{V}(t,x_{1},x_{2},x_{3}+L_{3})&=&\widetilde{V}(t,x_{1},x_{2},x_{3}).
\end{eqnarray}
In addition, one can verify that
\begin{equation}
\label{inticon}
\widetilde{V}(x_{0},x_{1},x_{2},0)=1.
\end{equation}
We see that $\widetilde{V}(x)$ is continuous on $R\otimes S^{1}\otimes S^{1}\otimes S^{1}$. Thus we can make the gauge transformation
\begin{equation}
A_{\mu}\to A_{\mu}^{\widetilde{V}}= \widetilde{V}(A_{\mu}+\frac{i}{g}\partial_{\mu})\widetilde{V}^{\dag},
\end{equation}
and have,
\begin{equation}
\label{gftr}
n\cdot A(x)^{\widetilde{V}}=\frac{1}{gL_{3}}\phi(x_{0},x_{1},x_{2},1)=\frac{1}{L_{3}}\int_{0}^{L_{3}}\ud z  n\cdot A^{\widetilde{V}}(x_{0},x_{1},x_{2},z),
\end{equation}
\begin{equation}
\label{wilgc}
e^{ig n\cdot A(x)^{\widetilde{V}}L_{3}}=\mathcal{P}\exp (\mathrm{i}g\int_{0}^{L_{3}}\ud z  n\cdot A(x_{0},x_{1},x_{2},z)).
\end{equation}
We see that one can choose a continuous gauge transformation so that
 \begin{equation}
 n\cdot\partial n\cdot A^{U}=0,\quad U(x_{0},x_{1},x_{2},0)=1.
 \end{equation}

We emphasize that the  traceless Hermitian matrix $\phi(x)$ is not necessarily to be continuous on the manifold $R^{4}$. That is to say, it may be impossible to choose a continuous traceless Hermitian matrix $\phi(x)$ so that
\begin{equation}
e^{\mathrm{i}\phi(x)}=\mathcal{P}\exp (\mathrm{i}g\int_{0}^{x_{3}}\ud z  n\cdot A(x_{0},x_{1},x_{2},z)),
\end{equation}
for some configurations of $A^{\mu}(x)$ and coupling constant $g$ even if one neglect the periodic boundary conditions. We do not consider such singularities here for simplicity, which will be discussed in other works.

\subsection{The degeneracy of the gauge fixing condition $n\cdot \partial n\cdot A=0$}

We consider the degeneracy of the gauge fixing condition (\ref{gaugex}) in this subsection.
We start from the gauge transformation of $n\cdot A(x)$,
\begin{equation}
n\cdot A(x)\to n\cdot A^{\prime}(x)\equiv U(x)(n\cdot A+\frac{i}{g}n\cdot\partial)U^{\dag},\quad U(x_{0},x_{1},x_{2},0)=1.
\end{equation}
If both $n\cdot A(x)$ and $n\cdot A^{\prime}(x)$ are independent of $n\cdot x$, then we have,
\begin{equation}
U(x)=e^{\mathrm{i}gn \cdot A^{\prime}(x)x_{3}}e^{-ign\cdot A(x)x_{3}}.
\end{equation}
Continuity of $U(x)$ on $R\otimes S^{1}\otimes S^{1}\otimes S^{1}$ requires that
\begin{eqnarray}
\label{eqgo}
e^{\mathrm{i}gn\cdot  A^{\prime}(x)L_{3}}&=&e^{\mathrm{i}gn\cdot A(x)L_{3}}.
\end{eqnarray}
Degeneracy of the gauge fixing condition (\ref{gaugex}) is caused by degeneracy of $n\cdot  A^{\prime}(x)$ with fixed $e^{\mathrm{i}gn\cdot  A^{\prime}(x)L_{3}}$ , which depends on the  value of $e^{\mathrm{i}gn\cdot  A^{\prime}(x)L_{3}}$.  According to (\ref{eqgo}), such degeneracy is the function of $e^{\mathrm{i}gn\cdot A(x)L_{3}}$. In addition,  one can choose a gauge transformation $U(x)$ so that
\begin{equation}
e^{\mathrm{i}g n\cdot A(x)^{U}L_{3}}=\mathcal{P}\exp (\mathrm{i}g\int_{0}^{L_{3}}\ud z  n\cdot A(x_{0},x_{1},x_{2},z))
,\quad U(x_{0},x_{1},x_{2},0)=1,
\end{equation}
as displayed in (\ref{inticon}), (\ref{gftr}) and (\ref{wilgc}). We thus have
\begin{equation}
\int[\mathcal{D}U] |Det(n\cdot \partial n\cdot D^{U})|\delta(n\cdot\partial n\cdot A^{U})
=N(\mathcal{P}\exp (\mathrm{i}g\int_{0}^{L_{3}}\ud z  n\cdot A(x_{0},x_{1},x_{2},z))).
\end{equation}
We further have
\begin{eqnarray}
1&=&\int[\mathcal{D}U]
\frac{|Det(n\cdot \partial n\cdot D^{U})|}
{N(\mathcal{P}\exp (\mathrm{i}g\int_{0}^{L_{3}}\ud z  n\cdot A(x_{0},x_{1},x_{2},z)))}
\delta(n\cdot\partial n\cdot A^{U})
\nonumber\\
&=&\int[\mathcal{D}U]
\frac{|Det(n\cdot \partial n\cdot D^{U})|}{N(e^{\mathrm{i}g n\cdot A^{U}L_{3}})}
\delta(n\cdot\partial n\cdot A^{U}).
\end{eqnarray}
According to the Faddeev-Popov procedure , we have
\begin{eqnarray}
\int[\mathcal{D}A]e^{\mathrm{i}\int\ud^{4}x\mathcal{L}(x)}&=&
\int[\mathcal{D}U]
\frac{|Det(n\cdot \partial n\cdot D^{U})|}{N(e^{\mathrm{i}g n\cdot A^{U}L_{3}})}
\delta(n\cdot\partial n\cdot A^{U})
\int[\mathcal{D}A]e^{\mathrm{i}\int\ud^{4}x\mathcal{L}(x)}
\nonumber\\
&=&
\int[\mathcal{D}A]e^{\mathrm{i}\int\ud^{4}x\mathcal{L}(x)}
\frac{|Det(n\cdot \partial n\cdot D )|}{N(e^{\mathrm{i}g n\cdot A L_{3}})}
\delta(n\cdot\partial n\cdot A),
\end{eqnarray}
where we have made use of the gauge invariance of the Lagrangian density.

We then consider the function $N(e^{\mathrm{i}g n\cdot A L_{3}})$.
We consider two Hermitian matrixes $ A_{1}$ and $A_{2}$,
\begin{equation}
A_{1}=U_{A_{1}}A_{1;\lambda}U_{A_{1}}^{\dag},\quad A_{2}=U_{A_{2}}A_{2;\lambda}U_{A_{2}}^{\dag},
\end{equation}
where $A_{i;\lambda}$($i=1,2$) are diagonal hermitian matrixes and $U_{A_{i}}$($i=1,2$) are unitary matrixes. Solution of the equation
\begin{equation}
e^{\mathrm{i}A_{1}}=e^{\mathrm{i}A_{2}}
\end{equation}
reads
\begin{equation}
\label{degenerator}
e^{\mathrm{i}(A_{1;\lambda}-A_{2;\lambda})}=1,
\end{equation}
\begin{equation}
 U_{A_{2}}^{\dag}U_{A_{1}}e^{\mathrm{i}A_{1;\lambda}}=e^{\mathrm{i}A_{1;\lambda}}U_{A_{2}}^{\dag}U_{A_{1}}.
\end{equation}
If eigenvalues of $e^{\mathrm{i}A_{1;\lambda}}$ are not degenerate then $U_{A_{2}}^{\dag}U_{A_{1}}$
is diagonal. We have,
\begin{equation}
U_{A_{1}}A_{1;\lambda}U_{A_{1}}^{\dag}=U_{A_{2}}A_{1;\lambda}U_{A_{2}}^{\dag}.
\end{equation}
Thus degeneracy of the gauge fixing condition $n\cdot \partial n\cdot A=0$ originates form the degeneracy of solutions of the equation (\ref{degenerator}), which is independent of $A^{\mu}(x)$,  once the matrix $e^{\mathrm{i}g n\cdot A L_{3}}$ is not degenerate.
We  see that $N(e^{\mathrm{i}g n\cdot A L_{3}})$  is independent of $A^{\mu}(x)$  in this case. If eigenvalues of $e^{\mathrm{i}g n\cdot A L_{3}}(x)$ are degenerate in area of which the $4$-dimensional volume equal to $0$, then the gauge transformation $U(x)$ can be determined according to the  continuity in the neighbourhood of these points. Thus $N(e^{\mathrm{i}g n\cdot A L_{3}})$ is independent of $n\cdot A$ unless eigenvalues of $e^{\mathrm{i}g n\cdot A L_{3}}(x)$ are degenerate in
area with nonzero $4$-dimensional volume. For the case that eigenvalues of $e^{\mathrm{i}g n\cdot A L_{3}}(x)$ are degenerate in such area, integral measure of these configurations equals to $0$. We thus have
\begin{equation}
N(e^{\mathrm{i}g n\cdot A L_{3}})=N,
\end{equation}
except for configurations that have zero measure.

According to above analyses, we have
\begin{equation}
\big<T\{\mathcal{O}\}\big>=\frac{\int[\mathcal{D}A^{\mu}(x)]
\frac{\mathcal{O} e^{\mathrm{i}\int\ud^{4}x\mathcal{L}(x)}}{N(e^{\mathrm{i}g n\cdot A L_{3}})}
\delta(n\cdot\partial n\cdot A)|Det(n\cdot \partial n\cdot D )|}
{\int[\mathcal{D}A^{\mu}(x)]\frac{e^{\mathrm{i}\int\ud^{4}x\mathcal{L}(x)}}{N(e^{\mathrm{i}g n\cdot A L_{3}})}
\delta(n\cdot\partial n\cdot A)|Det(n\cdot \partial n\cdot D )|},
\end{equation}
where $\mathcal{O}$ is  a gauge invariant operator. If the matrix element  is not divergent in the region in which eigenvalues of $N(e^{\mathrm{i}g n\cdot A})$ are degenerate in area with nonzero volume, then we have,
\begin{equation}
\label{toexv}
\big<T\{\mathcal{O}\}\big>=\frac{\int[\mathcal{D}A^{\mu}(x)]\mathcal{O}e^{\mathrm{i}\int\ud^{4}x\mathcal{L}(x)}
|Det(n\cdot \partial n\cdot D )|
\delta(n\cdot\partial n\cdot A)}
{\int[\mathcal{D}A^{\mu}(x)]e^{\mathrm{i}\int\ud^{4}x\mathcal{L}(x)}
|Det(n\cdot \partial n\cdot D )|
\delta(n\cdot\partial n\cdot A)}.
\end{equation}

\subsection{The determinant $Det(n\cdot \partial n\cdot D )$ and the ghost term}

We consider the determinant $Det(n\cdot \partial n\cdot D )$ in this subsection. For the case that $n\cdot A$ is independent of $n\cdot x$, we have,
\begin{equation}
Det(n\cdot \partial n\cdot D )=Det(\mathrm{i} n\cdot\partial)Det(-\mathrm{i}n\cdot D).
\end{equation}
The term $Det(\mathrm{i} n\cdot\partial)$ is independent of gauge potentials, which do not disturb us here.  Eigenvectors of the operator $-\mathrm{i}n\cdot D=-\mathrm{i}n\cdot \partial +gn\cdot A$ can be written as direct products of those of the operators $-\mathrm{i}n\cdot \partial$ and $n\cdot A$ as $[n\cdot \partial,n\cdot A ]=0$. We then have,
\begin{equation}
Det(-\mathrm{i}n\cdot D)=\prod_{\lambda}(\frac{2\pi n}{L_{3}}+g\lambda)=Det(-\mathrm{i}n\cdot D^{U}),
\end{equation}
where $\lambda$ represents eigenvalues of $n\cdot A$ and $U$ is a gauge transformation that is independent of $n\cdot x$.  We can rule out eigenvectors of $n\cdot\partial$ with zero eigenvalue as gauge transformations that do not rely on $n\cdot x$ can be eliminated by the gauge transformation $U$.  We notice that the matrix $n\cdot A_{bc}=\mathrm{i}n\cdot A^{a}f^{abc}$ is Hermitian and antisymmetric. If there is a vector $\epsilon^{a}$ which satisfy the equation
\begin{equation}
\mathrm{i}n\cdot A^{a}f^{abc}\epsilon^{c}=\lambda \epsilon^{b}
\end{equation}
with $\lambda$ a real number($\lambda\ne 0$), then we have
\begin{equation}
\mathrm{i}n\cdot A^{a}f^{abc}\epsilon^{\ast c}=(-\mathrm{i}n\cdot A^{a}f^{abc}\epsilon^{c})^{\ast}=-\lambda \epsilon^{\ast c}.
\end{equation}
We see that $-\lambda$($\lambda\ne 0$) is the eigenvalue of $n\cdot A$ once $\lambda$ is the eigenvalue of $n\cdot A$. They appear as pairs. We have
\begin{eqnarray}
Det(-\mathrm{i}n\cdot D)&=&\prod_{n,\lambda> 0}(\frac{4\pi^{2} n^{2}}{L_{3}^{2}}-g^{2}\lambda^{2})(\prod_{n>0} n^{2})^{N_{\lambda;0}}
\nonumber\\
&=&(\prod_{n>0} n^{2})^{N_{\lambda;0}}\prod_{n> 0,\lambda> 0}(\frac{4\pi^{2} n^{2}}{L_{3}^{2}}-g^{2}\lambda^{2})^{2},
\end{eqnarray}
where $N_{\lambda;0}$ represents the degeneracy of the operator $n\cdot A$ with zero eigenvalue. We thus have,
\begin{equation}
Det(-\mathrm{i}n\cdot D)>0.
\end{equation}

We can then write (\ref{toexv}) as
\begin{eqnarray}
\big<T\{\mathcal{O}\}\big>&=&\frac{\int[\mathcal{D}A^{\mu}(x)]\mathcal{O}e^{\mathrm{i}\int\ud^{4}x\mathcal{L}(x)}
|Det(\mathrm{i}n\cdot \partial)| Det(-\mathrm{i} n\cdot D )
\delta(n\cdot\partial n\cdot A)}
{\int[\mathcal{D}A^{\mu}(x)]e^{\mathrm{i}\int\ud^{4}x\mathcal{L}(x)}
|Det(\mathrm{i}n\cdot \partial)| Det(-\mathrm{i} n\cdot D )
\delta(n\cdot\partial n\cdot A)}
\nonumber\\
&=&
\frac{\int[\mathcal{D}A^{\mu}(x)]\mathcal{O}e^{\mathrm{i}\int\ud^{4}x\mathcal{L}(x)}
Det(\mathrm{i} n\cdot D )
\delta(n\cdot\partial n\cdot A)}
{\int[\mathcal{D}A^{\mu}(x)]e^{\mathrm{i}\int\ud^{4}x\mathcal{L}(x)}
Det(\mathrm{i} n\cdot D )
\delta(n\cdot\partial n\cdot A)}
\nonumber\\
&=&
\frac{\int[\mathcal{D}A^{\mu}(x)]\mathcal{O}e^{\mathrm{i}\int\ud^{4}x\mathcal{L}(x)}
Det(n\cdot \partial n\cdot D )
\delta(n\cdot\partial n\cdot A)}
{\int[\mathcal{D}A^{\mu}(x)]e^{\mathrm{i}\int\ud^{4}x\mathcal{L}(x)}
Det(n\cdot \partial n\cdot D )
\delta(n\cdot\partial n\cdot A)}.
\end{eqnarray}
We then bring in ghost fields and have
\begin{eqnarray}
\label{toexp}
\big<T\{\mathcal{O}\}\big>
&=&
\frac{\int[\mathcal{D}A^{\mu}(x)\mathcal{D}c(x)\mathcal{D}\bar{c}(x)]\mathcal{O}
e^{-\mathrm{i}\int\ud^{4}x(\frac{1}{2}tr[G_{\mu\nu}G^{\mu\nu}]+\bar{c}(n\cdot\partial n\cdot D)c )(x)}
\delta(n\cdot\partial n\cdot A)}
{\int[\mathcal{D}A^{\mu}(x)\mathcal{D}c(x)\mathcal{D}\bar{c}(x)]
e^{-\mathrm{i}\int\ud^{4}x(\frac{1}{2}tr[G_{\mu\nu}G^{\mu\nu}]+\bar{c}(n\cdot\partial n\cdot D)c )(x)}
\delta(n\cdot\partial n\cdot A)}.
\end{eqnarray}

\section{Propagators of Gluons in the Gauge $n\cdot\partial n\cdot A=0$}
\label{perturbationserie}

In this section, we consider propagators of gluons at tree level.  We rewrite (\ref{toexp}) as
\begin{eqnarray}
\big<T\{\mathcal{O}\}\big>
&=&
\lim_{\xi\to 0}\frac{\int[\mathcal{D}A^{\mu}(x)\mathcal{D}c(x)\mathcal{D}\bar{c}(x)]\mathcal{O}
e^{-\mathrm{i}\int\ud^{4}x(\frac{1}{2}tr[G_{\mu\nu}G^{\mu\nu}]+\bar{c}(n\cdot\partial n\cdot D)c )(x)}
e^{-\int\ud^{4}x \frac{tr[(n\cdot\partial n\cdot A)^{2}]}{\xi}}}
{\int[\mathcal{D}A^{\mu}(x)\mathcal{D}c(x)\mathcal{D}\bar{c}(x)]
e^{-\mathrm{i}\int\ud^{4}x(\frac{1}{2}tr[G_{\mu\nu}G^{\mu\nu}]+\bar{c}(n\cdot\partial n\cdot D)c )(x)}
e^{-\int\ud^{4}x \frac{tr[(n\cdot\partial n\cdot A)^{2}]}{\xi}}}.
\end{eqnarray}
We then have,
\begin{equation}
\int\ud^{4}x e^{ik\cdot x}T\big<A^{\mu}(x)A^{\nu}(0)\big>
=\lim_{\xi\to 0}\frac{-i}{k^{2}}(g^{\mu\nu}-\frac{n^{\mu}k^{\nu}+n^{\nu}k^{\mu}}{n\cdot k}+\frac{n^{2}k^{\mu}k^{\nu}}{(n\cdot k)^{2}}
+\frac{ik^{2}\xi k^{\mu}k^{\nu}}{(n\cdot k)^{4}})
\end{equation}
at tree level.
We see that  propagators of gluons are equivalent to those in the axial gauge for $n\cdot k\ne 0$.
The propagator is divergent for $n\cdot k=0$. This is because that the gauge fixing condition $n\cdot\partial n\cdot A=0$ does not rule out gauge transformations that are independent of $n\cdot x$.  Although such gauge transformation are ruled out in Sec.\ref{quantization} according to the constraint
\begin{equation}
 U(x_{0},x_{1},x_{2},0)=1,
\end{equation}
gauge transformations that vary slowly along the direction $n^{\mu}$ are not excluded by such constraint. Thus  modes with small $n\cdot k$ may cause divergences in the limit $n\cdot L\to \infty$ with $L^{\mu}=(0,L_{1},L_{2},L_{3})$.

To see the difference between contributions of  modes  with small $n\cdot k$  and those with $n\cdot k=0$. We first consider the quantity
\begin{equation}
\int\ud^{4}x e^{ik\cdot x}T\big<n\cdot A^{\mu}(x)n\cdot A^{\nu}(0)\big>.
\end{equation}
We have
\begin{equation}
\lim_{n\cdot k\to 0}\int\ud^{4}x e^{ik\cdot x}T\big<n\cdot A^{\mu}(x)n\cdot A^{\nu}(0)\big>=0
\end{equation}
at tree level. We then consider the mode
\begin{equation}
\tilde{A}=\frac{1}{n\cdot L}\int_{0}^{n\cdot L}\ud s n\cdot A(x^{\mu}+sn^{\mu}),
\end{equation}
where $n^{\mu}$ is the directional vector along $x_{i}$-axis($i=1,2,3$) and $n^{2}=-1$. We make the decomposition
\begin{equation}
A^{\mu}\equiv- \tilde{A}n^{\mu}+A^{\mu}+\tilde{A}n^{\mu}\equiv \tilde{A}n^{\mu}+\hat{A}^{\mu},
\quad \hat{A}^{\mu}\equiv A^{\mu}+\tilde{A}n^{\mu}
\end{equation}
and have
\begin{eqnarray}
tr[G^{\mu\nu}G_{\mu\nu}]&=&\frac{i}{g}tr[[\partial^{\mu}-ig\hat{A}^{\mu},\partial^{\nu}-ig\hat{A}^{\nu}
][\partial_{\mu}-ig\hat{A}_{\mu},\partial_{\nu}-ig\hat{A}_{\nu}]]
\nonumber\\
&&-2tr[\partial^{\mu} \tilde{A}\partial_{\mu}\tilde{A}]+O(g),
\end{eqnarray}
\begin{equation}
n\cdot \partial n\cdot A=n\cdot \partial n\cdot \hat{A}
\end{equation}
We further have
\begin{eqnarray}
\label{propn0}
&&\int\ud^{4}x e^{ik\cdot x}T\big<n\cdot A(x)n\cdot A(0)\big>
\nonumber\\
&=&
\int\ud^{4}x e^{ik\cdot x}T\big<\tilde{A}(x)\tilde{A}(0)\big>
+\int\ud^{4}x e^{ik\cdot x}T\big<n\cdot \hat{A}(x)n\cdot \hat{A}(0)\big>
\nonumber\\
&=&\frac{2\pi i\delta(n\cdot k)}{k^{2}}
\end{eqnarray}
at tree level. We see that propagators of gluons in the new gauge are different from those in the axil gauge for $n\cdot k=0$.

\section{Conclusions and Discussions}
\label{conc}

We present a gauge condition to quantize non-Abelian gauge theory on the manifold $R\otimes S^{1}\otimes S^{1}\otimes S^{1}$ in this paper, which reads $n\cdot \partial n\cdot A=0$.   We have proved that such gauge condition is continuous and free from the Gribov ambiguity for non-Abelian gauge theory considered here except for configurations with zero measure. We also prove that the functional determinant  $Det(-\mathrm{i}n\cdot D)$ is positive definite in the case that $n\cdot \partial n\cdot A=0$. Thus  ghosts can be brought in smoothly  even for one works in non-perturbative region.

Propagators of gluons in the new gauge are similar to those in the axial gauge at tree level for $n\cdot k\ne 0$. For the case that $n\cdot k=0$, propagators in the new gauge are different form those in the axial gauge as displayed in (\ref{propn0}). We notice that both the gauge condition $n\cdot\partial n\cdot A=0$  and the axial gauge $n\cdot A=0$ are equations of the field $n\cdot A$ and independent of other components of $A^{\mu}(x)$. Thus it is reasonable to believe that differences between these two gauge conditions will affect quantities involving $n\cdot A$.

\section*{Acknowledgments}
G. L. Zhou thanks Doctor Li-Ping Sun for helpful discussions and important suggestions on the manuscript. The work of G. L. Zhou is supported by The National Nature Science Foundation of China under Grant No. 11647022 and The Scientific Research Foundation for the Doctoral  Program of Xi'an University of Science and Technology under Grant No. 6310116055. The work of Z. X. Yan is supported by  The Department of Shanxi Province Natural Science Foundation of China under Grant No.2015JM1027.

\bibliography{gribov}

\end{document}